\documentclass[showpacs,preprintnumbers,amsmath,amssymb,pra]{revtex4}
\usepackage{graphicx}

\begin{document}

\def\br{{\bf r}}
\def\bp{{\bf p}}
\def\bv{{\bf v}}
\newcommand{\beq}{\begin{equation}}
\newcommand{\eeq}{\end{equation}}
\newcommand{\bea}{\begin{eqnarray}}
\newcommand{\eea}{\end{eqnarray}}

\title{Landau damping in trapped Bose-condensed gases}
\author{B. Jackson$^\dagger$ and E. Zaremba}
\affiliation{Department of Physics, Queen's University, Kingston, Ontario 
 K7L 3N6, Canada.}
\date{\today}
\begin{abstract}
We study Landau damping in dilute Bose-Einstein condensed gases in both
spherical and prolate ellipsoidal harmonic traps. We solve the 
Bogoliubov equations for the mode spectrum in both of these cases, and 
calculate the damping by summing over transitions between excited
quasiparticle states. The results for the spherical case are compared 
to those obtained in the Hartree-Fock approximation, where the 
excitations take on a single-particle character, and excellent
agreement between the two approaches is found.  We have also taken the 
semiclassical limit of the Hartree-Fock approximation and obtain
a novel expression for the Landau damping rate involving the time 
dependent self-diffusion function of the thermal cloud. As a final 
approach, we study the decay of a condensate mode by making use of 
dynamical simulations in which both the condensate and thermal cloud 
are evolved explicitly
as a function of time. A detailed comparison of all these methods 
over a wide range of sample sizes and trap geometries is presented. 
\end{abstract}
\maketitle

\section{Introduction}

One of the more challenging problems in the study of Bose-Einstein 
condensation (BEC) in trapped atomic gases concerns the finite 
temperature dynamics of a condensate interacting with a noncondensed, 
thermal component. Of particular interest in this regard is the study of
collective oscillations of the condensate, where one would expect to
see the influence of the thermal cloud on both the frequency and 
damping of the modes. Experiments which address these properties
have been performed in a number of laboratories around the 
world \cite{jin97,stamperkurn98,marago01,chevy02}, and provide
an ideal test-bed for the development of theories of Bose-condensed 
gases at finite temperatures.

One theoretical approach that can be used in the analysis of these 
experiments is the ZNG theory \cite{zaremba99}. In this
approach, the condensate is described by means of a generalized 
Gross-Pitaevskii equation while the dynamics of the thermal cloud is
determined by a Boltzmann-like kinetic equation. The full numerical
implementation of this theory has recently been presented
\cite{jackson02c} and its use in the analysis of a
variety of experiments has appeared in a series of 
papers \cite{jackson01,jackson02a,jackson02b,jackson02d}. 
An important part of this
previous work involved a detailed study of the collisional dynamics
which is responsible for the equilibration of the system with respect to
both energy and particle number. In this paper, however, we shall 
focus on an aspect for which collisions are of secondary importance, 
namely Landau damping.
This damping mechanism is dominant in the low density systems studied
experimentally~\cite{jin97,marago01,chevy02}). In this so-called
collisionless regime,
mean-field interactions mediate the transfer of energy from the 
condensate to the thermal component, leading to the damping of
condensate collective modes.

The subject of Landau damping in dilute BECs has been explored by several 
authors. Early work in an uniform gas was conducted by Hohenberg and Martin
\cite{hohenberg65} for low temperatures, while results for higher temperatures 
were obtained by Sz\'{e}pfalusy and Kondor \cite{szepfalusy74}. Liu 
\cite{liu97} used results obtained for the uniform gas to estimate the
damping in the case of a trapped gas while Pitaevskii and Stringari 
\cite{pitaevskii97} and Fedichev {\it et al.\ }\cite{fedichev98}
developed expressions for the Landau damping rate that explicitly
accounted for the trap geometry. Similar results 
were derived by Giorgini \cite{giorgini98,giorgini00} 
who also considered the Baliaev damping process whereby a collective
mode decays into two lower energy excitations. The latter can be
ignored, however, for the low-lying collective modes in a trapped gas
because of energy conservation restrictions.
Semiclassical results for Landau damping were 
also obtained using a dielectric formalism by Reidl 
{\it et al.\ }\cite{reidl00}, who found quite 
good agreement with the results of experiment \cite{jin97}.
The ZNG theory also provides a description of Landau damping and
applications of the theory to a number of different condensate
collective modes~\cite{jackson01,jackson02a,jackson02b} 
found damping rates that agreed well with experiment.

The perturbation theory approach of Refs.\ 
\cite{pitaevskii97,giorgini98,giorgini00} provides an expression for the
Landau damping rate that involves a sum over transitions between pairs 
of Bogoliubov excitations. The first fully quantum mechanical
evaluation of this expression was performed by Guilleumas and 
Pitaevskii (which we shall hence forward denote as G-P I) 
\cite{guilleumas00}, for a spherically-symmetric trapping potential. 
We should also mention that a related calculation was
performed by Das and Bergeman \cite{das01}, obtaining similar results to
G-P I. In the ZNG theory, the description of Landau damping is quite
different since it involves the dynamical evolution of the condensate in
the presence of a thermal cloud treated semiclassically at the level
of the Hartree-Fock approximation. The relation of this
approach to that used in G-P I is far from evident, although it is clear
that the physical basis of the two is the same. To investigate this
question, we performed numerical simulations in an earlier paper
\cite{jackson02c} for the system studied in G-P I and in fact found
quantitative agreement. One of the aims of the present paper is to 
compare the two methods over a much larger range of condensate sizes, 
which means that we have had to repeat the kind of calculations
performed in G-P I. By doing so, we are then able to
investigate the consequences of using Hartree-Fock, as opposed to
Bogoliubov, excitations in the Landau damping calculation. A
reformulation in terms of Hartree-Fock excitations has the added
benefit of allowing the semiclassical limit to be taken in a
straightforward way, thereby completing the connection to
the ZNG theory.

Perhaps an even more interesting question concerns the damping rate in
anisotropic traps. This issue arose in experiments on the transverse 
breathing mode in a highly elongated trapped gas~\cite{chevy02},
where the damping rate was found to be anomalously low -- around an 
order of magnitude smaller than one would expect in similar
circumstances on the basis of both experiment and theory. In Ref.\ 
\cite{jackson02b} we simulated this experiment and found that the low 
damping rate could be explained as a consequence of an in-phase 
collective oscillation of both the condensate and thermal cloud. 
On the other hand, if the thermal cloud is not set into oscillation,
the Landau damping rate is found to have a value typical of that
seen in other experiments. This result apparently contradicts the
conclusion of Guilleumas and Pitaevksii (G-P II) \cite{guilleumas02} who
performed a Landau damping calculation for an infinitely long
cylindrical condensate. They found a damping rate which was extremely 
low, around an order of magnitude smaller than the experimental results,
and {\it two} orders of magnitude smaller than our simulation results
for a stationary thermal cloud. A natural question concerns the validity
of modelling the experimental geometry by an infinite cylindrical trap
which changes the nature of the excitation spectrum and presumably the
damping rate. In this paper we address this question by repeating the
G-P I calculations as a function of the trap anisotropy and compare
these results with those obtained from our simulations. Our tentative
conclusion is that the damping rate has an anomalous dependence on
anisotropy in the limit of the infinite cylinder, so while the G-P II 
result is correct, it does not apply at the experimental anisotropy.
\section{Theory}

\subsection{Landau damping of Bogoliubov excitations}

For purposes of completeness it is useful to summarize the Bogoliubov 
theory and
the way in which a condensate collective mode is damped by thermal
excitations. In a second-quantized notation, the Hamiltonian is given by
\beq
\hat H = \int d\br\, \hat \psi^\dagger (\br) \left ( -{\hbar^2 \nabla^2
\over 2m} + V_{\rm ext}(\br) \right ) \hat \psi (\br) +{g\over 2} \int
d\br\, \hat \psi^\dagger (\br) \hat \psi^\dagger (\br) \hat \psi
(\br) \hat \psi (\br)
\end{equation}
where $V_{\rm ext}(\br)$ is the trapping potential and $g = 4\pi a
\hbar^2/m$ is the interaction parameter defined in terms of the $s$-wave
scattering length $a$. If the field operator is expressed as
\beq
 \hat \psi = \Phi_0 + \tilde \psi,
 \eeq
where $\Phi_0$ is the condensate wavefunction and $\tilde \psi$ is the
excitation (or fluctuation) field operator, the Bogoliubov approximation
is generated by expanding $\hat H$ to second order in $
\tilde \psi$. The terms linear in $\tilde \psi$ vanish if $\Phi_0$
satisfies the Gross-Pitaevskii equation
\beq
\left ({-\hbar^2 \nabla^2 \over 2m}+ V_{\rm ext} + g n_{c0} \right )
\Phi_0 = \mu \Phi_0\,,
\eeq
where $n_{c0} = |\Phi_0|^2$ is the condensate density. At this level of
approximation, the condensate does not interact directly with the 
thermal excitations as determined in the Bogoliubov theory.

The resulting Bogoliubov Hamiltonian is then diagonalized by means of 
the Bogoliubov transformation
\beq
 \tilde \psi(\br) = \sum_i \left (u_i(\br) \alpha_i - v_i^*(\br)
 \alpha_i^\dagger \right )\,.
\eeq
The quasiparticle amplitudes $u_i$ and $v_i$ satisfy the Bogoliubov
equations
\begin{eqnarray}
\hat L u_i - g n_{c0} v_i &=& E_i u_i\nonumber \\
\hat L v_i - g n_{c0} u_i &=& -E_i v_i\,,
\label{Bogol}
\end{eqnarray}
where we introduce the operator $\hat L\equiv -\hbar^2 \nabla^2/2m + 
V_{\rm ext}+2gn_{c0}-\mu$. The orthonormality of the quasiparticle
amplitudes is specified by the relation
\beq
\int d\br\, \left [ u^*_i(\br) u_j(\br) - v^*_i(\br) v_j(\br) \right ] =
\delta_{ij}\,.
\label{orthonorm}
\eeq
Apart from a constant, the Bogoliubov Hamiltonian is given by
\beq
\hat H_B = \sum_i E_i \alpha_i^\dagger \alpha_i\,,
\eeq
where the excitation energies $E_i$ correspond to condensate collective 
modes. In thermal equilibrium, these modes are populated according to a 
Bose distribution $f_0(E_i)=[\exp(\beta E_i)-1]^{-1}$ with $\beta 
= (k_B T)^{-1}$.

The cubic terms in the expansion of $\hat H$,
\beq
V^{(3)} = g\int d\br\, \Phi_0\left (\tilde \psi^\dagger \tilde \psi \tilde \psi
+ \tilde \psi^\dagger \tilde \psi^\dagger \tilde \psi \right )\,,
\eeq
couple the Bogoliubov excitations and lead to a mechanism for
their decay. In particular, a mode which has been excited, and 
therefore no longer in equilibrium with the other thermal excitations,
will experience a decay known as Landau damping. To represent this
situation, one defines a state in which the mode occupation $n_{\rm
osc}$ is large compared to the equilibrium value. The energy in this
mode, $E_{\rm osc} = \hbar \omega_{\rm osc} n_{\rm osc}$, then decays as
$n_{\rm osc}$ goes to equilibrium. The transition rate from 
the initial nonequilibrium
state to any other state can be determined by perturbation theory 
\cite{pitaevskii97} and
the average rate of change of the energy in this mode is found to be
\beq
{\dot E_{\rm osc}\over E_{\rm osc}} = -{2\pi \over \hbar} \sum_{ij} |A_{ij}|^2 
(f_i -f_j) \delta(E_j - E_i -  \hbar \omega_{\rm osc})\,,
\eeq
where the transition matrix element is
\beq
A_{ij} = 2g \int d\br\,\Phi_0\left [ \left (u_i u_j^* -v_i u_j^* +v_i
v_j^* \right ) u_{\rm osc} - \left ( u_i u_j^* -u_i v_j^* + v_i v_j^*
\right ) v_{\rm osc} \right ]\,.
\label{matrix}
\eeq
For future reference we note that the Hartree-Fock approximation is 
recovered by setting $v_i$ (but {\it not} $v_{\rm osc}$) in 
(\ref{matrix}) equal to
zero and determining $u_i$ from the first of the Bogoliubov equations in
(\ref{Bogol}) with $v_i=0$.

The damping rate $\Gamma$ is defined according to $2\Gamma = - \dot
E_{\rm osc}/E_{\rm osc}$ (which implies that the {\it amplitude} of the
mode decays as $e^{-\Gamma t}$). Thus,
\beq
\Gamma = {\pi \over \hbar} \sum_{ij} |A_{ij}|^2 (f_i -f_j)
\delta(E_j - E_i -  \hbar \omega_{\rm osc})\,.
\label{dampingrate}
\eeq
This expression is the basis of the calculation of Landau damping as
carried out by Guilleumas and Pitaevskii for both spherically symmetric
and cylindrical traps \cite{guilleumas00,guilleumas02}. In order 
to display the results of our calculations it is convenient to follow
their notation
\beq
\frac{\Gamma}{\omega_{\rm osc}}=\sum_{ij} \gamma_{ij} \delta (\omega_{ij}
 -\omega_{\rm osc}),
\label{dampingrate2}
\eeq
where
\beq
\gamma_{ij}=\frac{\pi}{\hbar^2 \omega_{\rm osc}}
 |A_{ij}|^2 (f_i - f_j),
\label{dampingstrength}
\eeq
is a ``damping strength'' associated with
each possible transition, with frequency difference
$\omega_{ij}=(E_j-E_i)/\hbar$. 

The calculation of the damping then consists of solving the Bogoliubov
equations (\ref{Bogol}) for $u_{\rm osc}$, $v_{\rm osc}$ and
$\omega_{\rm osc}$ for the classical
mode of interest, together with the corresponding quantities for the 
thermally-populated excitations. The matrix elements $A_{ij}$ can then
be evaluated for transitions between each pair of excitations, which in 
turn yields $\gamma_{ij}$. More details of these calculations will be 
given in the next section, while the calculation of the total damping
rate $\Gamma$, including how we deal with the delta function in 
(\ref{dampingrate2}), will be presented in Section III. 

\subsection{Anisotropic traps}

In performing the above calculations, a significant saving in 
computational effort can be realised by considering a spherically 
symmetric trap. Not only does (\ref{matrix}) effectively reduce to a 
one-dimensional integral, but the spherical symmetry also reduces the
number of excitations that have to be considered. In
particular, if we label the modes with the usual quantum numbers 
$\{n,l,m\}$, then states corresponding to a given $l$ are $(2l+1)$-fold
degenerate, and $A_{ij}$ need only be calculated explicitly for one of
these states. This spherical case was first studied in Ref.\ 
\cite{guilleumas00}, and we will repeat some of these calculations in 
Section 3A in order to compare various approximations. However, here
we are also interested in the damping in anisotropic traps which is
the most common experimental situation. It turns out that
calculations for anisotropic traps, even those with axial symmetry, are
much more involved since the degeneracy of excitations with different 
$m$ quantum numbers is lifted. As a result, determining the
Bogoliubov excitation spectrum requires the diagonalization
of much larger matrices than in the spherical case. The corresponding 
increase in the number of possible transitions,
along with the fact that evaluation of the $A_{ij}$ matrix elements 
now involves integrating over both radial and angular variables,
significantly increases computational overheads. In the remaining part
of this section we shall describe how we tackle these problems by making
use of a spherical harmonic basis. The calculation for an isotropic trap 
is then a simplified version of this general scheme.
 
As a first step, we find the ground state condensate wavefunction
as detailed in Ref.\ \cite{hutchinson98}. This makes use
of a set of normalised basis functions $\psi_{nlm} (\br) = R_{nl} (r) 
Y_{lm} (\theta,\phi)$, which are eigenfunctions of the Hamiltonian 
$\hat{h}_0 = -\hbar^2\nabla^2/2m+V_0 (r)$. The potential $V_0 (r)$ is 
the $l=0$ component of the total axially-symmetric effective potential 
$V(\br) 
\equiv V_{\rm ext}(\br)+gn_c(\br)$ when expanded in terms of Legendre 
polynomials, $V (r,\theta)=\sum_l V_l (r) P_l (\cos \theta)$. We take the
external harmonic potential to be of the form $V_{\rm ext}(\br) =
(m\omega_\perp^2/2)(x^2+y^2+\lambda^2 z^2)$, where $\lambda$ is the
anisotropy parameter. By defining
a nonspherical perturbation $\Delta V(\br)=V(\br)-V_0(r)$, and 
expanding an 
arbitrary solution of the GP equation as $\phi(\br)=\sum_{nl} a_{nl} 
\psi_{nlm}(\br)$, one can set up a matrix equation for the coefficients 
$a_{nl}$. The condensate wavefunction is then given by the lowest-energy
even-parity solution  $\Phi_0(\br)=\sqrt{N_c} \phi_0(\br)$, which in 
turn can be used to calculate the condensate density $n_c(\br)$. 
Updating the effective potential $V(\br)$ then leads to a new matrix 
equation, which is solved to yield an updated wavefunction. The 
calculation is repeated until $\Phi_0$ converges
to the correct condensate wavefunction, with the lowest eigenvalue
giving the chemical potential $\mu$. In the process, one also 
generates a full set of eigenfunctions $\phi_\alpha(\br)$
for the Hamiltonian $\hat{h} \equiv -\hbar^2 \nabla^2/2m + 
V(\br)-\mu$ with eigenvalues $\varepsilon_\alpha$.

Once we have this information, the Bogoliubov equations can be solved by
introducing the expansion $\psi_i^{+}(\br)=\sum_{\alpha} 
c_{\alpha}^{(i)} \phi_{\alpha}(\br)$, where 
$\psi_i^{\pm}(\br)=u_i(\br)\pm v_i(\br)$. The index $i$ labels the
different excitations and includes the azimuthal quantum number $m$.
Again, this leads to a matrix equation, which can be 
solved
to obtain the eigenfunctions $\psi_i^{\pm}(\br)$ and energy 
eigenvalue $E_i$ for each mode. For the purposes of the Landau damping 
calculation, which we move onto next, it is convenient to express these
functions as
\begin{eqnarray}
\label{eq:psi+}
 \psi_i^{+}(\br)&=&\sum_{nl}d_{nl}^{(i)}R_{nl}(r)Y_{lm}(\theta,\phi), \\
\label{eq:psi-}
 \psi_i^{-}(\br)&=&\sum_{nl}e_{nl}^{(i)}R_{nl}(r)Y_{lm}(\theta,\phi),
\end{eqnarray}
by making use of the expansions of $\phi_\alpha$ in terms of the
$\psi_{nlm}$.

For the particular mode of interest (i.e.,\ the one for which we are calculating the 
damping) we can define the mode densities, $\delta n^+(\br)=\Phi_0(\br) 
\psi_{\rm osc}^{+}(\br)$, and
$\delta n^-(\br)=\Phi_0(\br)\psi_{\rm osc}^{-}(\br)$. Using this 
notation, the Landau damping matrix elements become
\beq
A_{ij} = \frac{g}{2} \int d\br\, \left [ \delta n^- (\psi_j^{+*} 
\psi_i^{+} + 3 \psi_j^{-*} \psi_i^{-})
 + \delta n^+ (\psi_j^{+*} \psi_i^{-} - 
 \psi_j^{-*} \psi_i^{+})\right ]\,.
\label{eq:landau-ani}
\eeq
%
In evaluating this integral, it is convenient to expand the mode densities 
(where the mode has a particular $m$-number, which we denote $\bar{m}$)
in terms of spherical harmonics
\begin{eqnarray}
 \label{eq:deln+}
 \delta n^\pm (\br) &=& \sum_{\bar{l}} \sqrt{\frac{4\pi}{2\bar{l}+1}}
 \delta n_{\bar{l}}^\pm (r) Y_{\bar{l}\bar{m}}(\theta,\phi)\,,
\end{eqnarray}
where
\begin{equation}
 \delta n_{l}^{\pm} (r) = \sqrt{\frac{2l+1}{4\pi}} \int^{2\pi}_0 d\phi 
 \int^{\pi}_0 d\theta \sin \theta Y_{l{\bar m}}^* (\theta,\phi) \delta n^{\pm} (\br).
\end{equation}
Substituting  (\ref{eq:psi+}), (\ref{eq:psi-}) and (\ref{eq:deln+})
into (\ref{eq:landau-ani}) gives
\begin{eqnarray}
 A_{ij}&=&\frac{g}{2} \int^{\infty}_0 dr r^2 \sum_{nln'l'} R_{nl}(r) 
 R_{n'l'}(r) \left [ \left (d_{nl}^{(j)} d_{n'l'}^{(i)} +
 3e_{nl}^{(j)}e_{n'l'}^{(i)} \right ) B_{ll'ij}^- (r) \right .
 \nonumber \\
&& \hskip 2truein + \left . \left (d_{nl}^{(j)} e_{n'l'}^{(i)} - 
e_{nl}^{(j)} d_{n'l'}^{(i)} \right ) B_{ll'ij}^+(r) \right ],
\label{eq:laudamp-asy}
\end{eqnarray}
where
\begin{equation}
 B_{ll'ij}^{\pm} (r) = \sqrt{\frac{2l'+1}{2l+1}} \sum_{\bar{l}} \langle \bar{l}
 l' 0 0 | l 0 \rangle \langle \bar{l} l' \bar{m} m_i | l m_j \rangle
 \delta n_{\bar{l}}^{\pm} (r),
\end{equation}
in which $\langle l_1 l_2 m_1 m_2 | l_3 m_3 \rangle$ is the
Clebsch-Gordan coefficient \cite{messiah66}, and where $m_j=m_i+\bar{m}$. The 
calculation of the Landau damping then involves
evaluation of (\ref{eq:laudamp-asy}) for each pair of excitations with a 
difference of energies in a particular interval $E_{\rm min}<E_j-E_i <
E_{\rm max}$. In this paper we shall focus on the Landau damping of modes
with $\bar{m}=0$ and even parity, so that selection rules connect excitations
with $\Delta m=0$ and the same parity. Unlike the spherically-symmetric
case, however, excitations with different $m$ values must be summed
explicitly.
Nevertheless, since the contributions from the $\pm m_i$ excitations
are identical, a two-fold saving can be gained
for $m_i \neq 0$. Our results for 
the breathing mode as a function of anisotropy will be presented in 
Section 3B. 

\subsection{Hartree-Fock Approximation}
\label{HFA}

In this section we obtain an expression for Landau damping in the
Hartree-Fock (HF)
approximation. Although this can be done trivially by setting $v_i =0$
in (\ref{matrix}), it is useful to rederive the result using an 
alternative method for a number of reasons. First it clearly shows that
Landau damping is associated with the work done on
the thermal cloud by the
dynamic mean field of the oscillating condensate. Second, this
reformulation allows one to take the semiclassical limit in a
straightforward way, thereby facilitating a comparison with the
semiclassical simulations that we have performed on the basis of the ZNG
theory.

Within the HF approximation, the condensate and thermal cloud are
treated as two distinct components. The condensate evolves
dynamically according to a time-dependent GP equation while the thermal
cloud responds to the condensate mean field. This picture is quite
distinct from the Bogoliubov approach in which the `condensate mode'
is distinguished from the `thermal excitations' only through the 
assumed different occupations of the respective modes. Common to both
pictures, however, is the absence of a thermal cloud mean field acting
back on the condensate. Such effects are included in the ZNG theory and
are crucial for the satisfaction of the generalized Kohn
theorem~\cite{zaremba99}.

For small
deviations from equilibrium, the time-dependent condensate
wavefunction is given by
\begin{equation}
\Phi(\br,t) = \Phi_0(\br)+\delta \Phi(\br,t)\,,
\end{equation}
where the fluctuation $\delta\Phi$ is obtained from the linearized GP
equation. The fluctuating part of the condensate density is then
\begin{equation}
\delta n_c(\br,t) = \Phi_0(\br) \left [ \delta \Phi(\br,t) + \delta
\Phi^*(\br,t) \right ]
\end{equation}
which gives rise to a dynamic mean field $2g\delta n_c(\br,t)$ acting on
the thermal cloud. Thus the thermal cloud experiences a perturbation
\begin{equation}
H'(t) = \int d\br\, 2g \delta n_c(\br,t) \hat n(\br)\,,
\label{perturbation}
\end{equation}
where $\hat n(\br)$ represents the density operator of the thermal
component. The linearized GP equation has solutions of the form
\begin{equation}
\delta \Phi(\br,t) = u_{\rm osc}(\br) e^{-i\omega_{\rm osc} t} - v_{\rm osc}^*(\br) 
 e^{i\omega_{\rm osc} t}\,,
\end{equation}
where the $u$ and $v$ are, apart from a normalization, the same
Bogoliubov amplitudes introduced earlier and $\omega_{\rm osc}$ is the 
frequency of the particular condensate mode of interest. In terms of 
this wavefunction, the condensate density fluctuation is given by
\bea
\delta n_c(\br,t) &=& \Phi_0(\br) \psi_{\rm osc}^{-}(\br)
e^{-i\omega_{\rm osc} t} + 
{\rm c.c.} \nonumber \\
&\equiv& \delta n^-(\br)  e^{-i\omega_{\rm osc} t} + {\rm c.c.}
\eea
where
\begin{equation}
\psi_{\rm osc}^{-}(\br) = u_{\rm osc}(\br) - v_{\rm osc}(\br)\,.
\end{equation}

With this form of the condensate density fluctuation, the perturbation 
given in (\ref{perturbation}) has a harmonic time dependence and can 
be treated by means of time-dependent perturbation
theory. One thus finds that the time-averaged rate of
change of the thermal cloud energy is given to lowest order in the
perturbation by the expression
\begin{equation}
\overline{{dE \over dt}} = 2\omega_{\rm osc} (2g)^2 \int d\br \int d\br'\,
\delta n^{-*}(\br) \chi^{\prime\prime}(\br,\br',\omega_{\rm osc})
\delta n^{-}(\br')
\label{rate}
\end{equation}
where $\chi^{\prime\prime}(\br,\br',\omega)$ is the imaginary part of the 
time Fourier transform of the thermal cloud density response function
\begin{equation}
\chi(\br,\br',t-t') = {i\over \hbar} \theta(t-t')\langle [\hat n(\br,t),
\hat n(\br',t') ] \rangle_0 \,.
\label{response}
\end{equation}
The angular brackets, $\langle ... \rangle_0$, denote an equilibrium
expectation value. Treating the thermal atoms as independent HF
excitations, the spectral density is given by the expression
\begin{equation}
\chi^{\prime\prime}(\br,\br',\omega) = \pi \sum_{ij} \phi_i^*(\br) \phi_j(\br)
\phi_j^*(\br') \phi_i(\br') [f_i -f_j] \delta(\varepsilon_j - \varepsilon_i
- \hbar \omega)
\end{equation}
where $\phi_i(\br)$ is the HF wavefunction (as obtained from
(\ref{Bogol}) by setting $v_i =0$)  for the $i$-th state with energy
$\varepsilon_i$, and $f_i$ is the thermal Bose occupation number.
Substituting this expression into (\ref{rate}), we find
\begin{equation}
\overline{{dE \over dt}} = 2\pi \omega_{\rm osc} \sum_{ij}|A_{ij}|^2(f_i-f_j)
\delta(\varepsilon_j - \varepsilon_i - \hbar \omega_{\rm osc})
\label{rate2}
\end{equation}
with
\begin{equation}
A_{ij} = 2g \int d\br\,\phi_i(\br) \phi_j^*(\br)
\delta n^{-}(\br)\,.
\end{equation}
As stated earlier, this matrix element can be obtained from 
(\ref{matrix}) by simply setting $v_i=0$, but retaining $v_{\rm osc}$ 
which is needed to
properly define the condensate density fluctuation. 

The rate of change of the condensate energy $\dot E_{\rm osc}$ is of 
course the negative of (\ref{rate2}). To extract a damping rate we 
must divide
this by the energy of the mode which is determined by the amplitude of
the condensate density fluctuation. If $u$ and $v$ are normalised 
according to (\ref{orthonorm}), this energy is just $\hbar \omega_{\rm
osc}$.

\subsection{Semiclassical HF}
\label{SCHF}

Our purpose in this section is to show how the quantal HF damping can be
reduced to a semiclassical form. This will allow us to make contact
with the semiclassical ZNG simulations. The response function in 
(\ref{response}) can 
be written $\chi(\br,\br',t-t') = 2i\theta(t-t')\phi(\br,\br',t-t')$ 
with
\begin{equation}
\phi(\br,\br',t-t') = {1\over 2\hbar} \langle [\hat n(\br,t),
\hat n(\br',t') ] \rangle_0\,.
\end{equation}
This correlation function can be expressed in the form~\cite{kubo57}
\begin{equation}
\phi(\br,\br',t) = {i\over 2} \int_0^\beta d\lambda \langle 
\hat {\dot n}(\br,t-i\hbar \lambda) \hat n(\br') \rangle_0
\end{equation}
which allows the semiclassically limit for the dynamics to be taken 
straightforwardly. Setting $\hbar \to 0$, we have
\begin{equation}
\phi(\br,\br',t) = {i\beta \over 2} \langle 
\dot n(\br,t) n(\br') \rangle_0
\label{semiclass}
\end{equation}
where now $n(\br,t)$ is a classical density variable.
The Fourier transform of (\ref{semiclass}) gives
\begin{equation}
\phi(\br,\br',\omega) = {\beta \omega \over 2} \int dt\, e^{i\omega t}
\langle n(\br,t) n(\br') \rangle_0\,.
\label{correlation}
\end{equation}

To evaluate the correlation function in (\ref{correlation}), we 
consider a thermal cloud containing $\tilde N$ atoms. The density of 
the cloud at time $t$ is thus given by
\beq
n(\br,t) = \sum_{i=1}^{\tilde N} \delta(\br - \br_i(t))
\eeq
where $\br_i(t)$ is the position of the $i$-th particle at time $t$.
Thus,
\beq
\langle n(\br,t) n(\br') \rangle_0 = \sum_{ij} \langle \delta(\br -
\br_i(t)) \delta(\br' - \br_j(0)) \rangle_0\,.
\eeq
Since the atoms in the cloud are independent and equivalent, we have
\beq
\langle n(\br,t) n(\br') \rangle_0 = G_s(\br,\br',t) + \tilde n_0(\br)
\tilde n_0(\br')
\eeq
where we have defined the self-diffusion function~\cite{vanhove54}
\beq
G_s(\br,\br',t) = \tilde N \langle \delta(\br - \br_1(t)) \delta(\br' -
\br_1(0)) \rangle_0
\eeq
and $\tilde n_0(\br)$ is the equilibrium thermal cloud density. Since
the product of equilibrium densities is time-independent, it will not
contribute to the Fourier transform in (\ref{correlation}) at finite 
$\omega$ and can be
dropped. Combining these results, the semiclassical (sc) Landau damping 
rate is given by 
\beq
{\Gamma_{{\rm sc}} \over \omega_{\rm osc}} = {2g^2 \over \hbar k_B T} 
\tilde f(\omega_{\rm osc})
\label{gamma_sc}
\eeq
where $\tilde f(\omega)$ is the Fourier transform of the function
\beq
f(t) = 
\int d\br \int d\br' \delta n^{-*}(\br) 
G_s(\br,\br',t) \delta n^{-}(\br')\,.
\label{f(t)}
\eeq
We thus see that the quantity determining the damping rate is an
appropriately weighted spatial average of the self-diffusion function 
$G_s$. The latter can be evaluated analytically for some simple model
situations which provide insight into its behaviour. However in the
present situation, the thermal atoms are moving in the combined
potential of the trap and the mean field of the condensate and $G_s$ can
only be determined numerically. It is then
easier to deal with $f(t)$ directly. Using the definition of 
$G_s$, $f(t)$ can be expressed as
\beq
f(t) = \tilde N \langle  \delta n^{-*}(\br_1(t))  \delta
n^{-}(\br_1(0)) \rangle_0\,,
\eeq
which is convenient for numerical evaluation.

We note that $\br_1(t)$ is the
position of an atom at time $t$ starting from some initial position
$\br_1(0) \equiv \br_0$ with momentum $\bp_1(0) \equiv \bp_0$. The
initial phase point is distributed according to the equilibrium Bose
distribution $f_0(\br_0,\bp_0)$, normalized to $\tilde N$, and so the
correlation function takes the form
\beq 
f(t) = \int {d\br_0 d\bp_0 \over h^3} f_0(\br_0,\bp_0)  \delta
n^{-*}(\br_1(\br_0,\bp_0;t))  \delta n^{-}(\br_0)
\eeq
Thus, $f(t)$ can be determined by averaging over all possible
trajectories starting from some initial phase point. In practice this can
be done by considering an ensemble of $N_T$ test particles distributed 
in phase space according to $f_0(\br_0,\bp_0)$ and then performing the
discrete sum
\beq
f(t) \simeq {\tilde N \over N_T} 
\sum_{i=1}^{N_T} \delta n^{-*}(\br_1(\br_i,\bp_i;t))  \delta
n^{-}(\br_i)
\label{sum}
\eeq
$N_T$ is chosen sufficiently large to ensure that statistical
fluctuations in $f(t)$ are acceptably small. Examples of such
calculations will be given later. Finally, a numerical Fourier
transform of $f(t)$ at $\omega = \omega_{\rm osc}$ yields the 
semi-classical damping rate.

The value of $f(0)$ can also be evaluated directly as
\beq
f(0) = \int d\br\,\tilde n_0(\br) |\delta n^{-}(\br)|^2\,.
\eeq
This serves as a check of the sum over test particles described above.
In addition, we note that $f(t)$ tends to a finite limiting value for $t
\to \infty$. This limiting value can be subtracted from $f(t)$ in the
evaluation of $\tilde f(\omega)$ since this only affects  $\tilde
f(\omega)$ at $\omega =0$.

\subsection{ZNG Simulations}
\label{simulations}

Our final method for evaluating the Landau damping is based on $N$-body 
simulations~\cite{jackson02c} in which collisions are neglected. In 
this numerical scheme, the condensate wavefunction is evolved in time
by numerically solving the generalized GP equation using a split-operator FFT 
method, while the evolution of the thermal cloud is calculated by
sampling with a large number of test particles, which are evolved classically.
This is in fact equivalent to solving a collisionless Boltzmann 
kinetic equation for the thermal cloud, where the thermal excitations are 
described according to the semiclassical Hartree-Fock (HF) approximation 
(i.e., as
single particle excitations in a quasi-uniform gas \cite{zaremba99}). 
The dynamic mean field of the thermal cloud then gives rise to decay 
of the condensate oscillation, which is identified with 
Landau damping. Full details of the numerical methods are given in 
Ref.\ \cite{jackson02c}. 

The first step in the calculation is the self-consistent determination
of the condensate wavefunction together with the thermal cloud density.
The simulation is then initiated by exciting the condensate in such a 
way as to realise the relevant mode of study. It is straightforward to 
show that this can be achieved by adding $\psi_{\rm osc}^{-} (\br)$, 
as obtained from the solution of the Bogoliubov equations (\ref{Bogol}),
to the condensate ground state wavefunction. The 
subsequent evolution of the mode of interest 
can then be monitored by evaluating the moments $\langle x^2\rangle $, 
$\langle y^2\rangle $, and $\langle z^2\rangle $ of the condensate as a 
function of time. These are calculated numerically using 
$\langle \chi\rangle (t)= \int d \br \,\chi |\Phi (\br,t)|^2$. 
A plot against time yields a damped oscillation, which is 
fit to a sinusoidal function with an exponentially
decaying factor $\exp(-\Gamma t)$. Each simulation extends for a time
of $\omega_\perp t = 10$, where the relatively short time is chosen so
as to avoid driving the thermal cloud too far out of equilibrium.
This point is discussed more fully in 
\cite{jackson02c}. The damping rate, $\Gamma$, is then
extracted from the simulation data, which can then be compared to the results 
of our other calculations.  It should be emphasized that within this
approach the damping of the condensate oscillation is a direct result of
the dynamic mean field that the thermal cloud exerts on the condensate.
In other words, no damping would occur if the condensate were to simply
oscillate in the static equilibrium mean field of the thermal cloud.

\section{Results}

\subsection{Isotropic Traps}

\begin{figure}[here]
\scalebox{0.45}
{\includegraphics{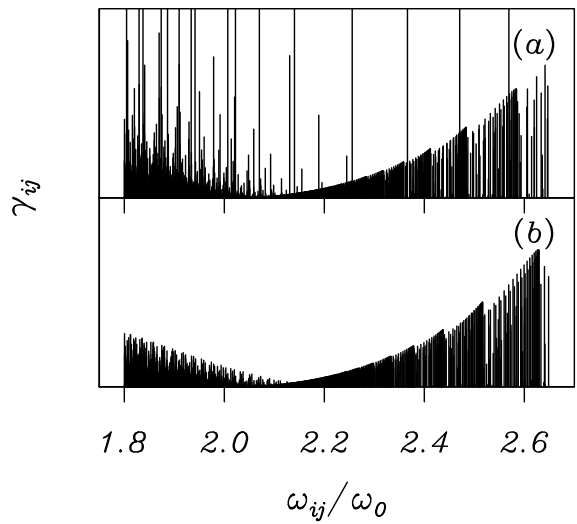}}

\caption{\label{fig:land-hist}
 Histogram showing results of Landau damping calculations for
 $N_c=2.5\times 10^5$ $^{87}$Rb atoms at a temperature of $k_B T/\mu=1.5$.
 The height of the bars represents the damping strength $\gamma_{ij}$ of each
 transition, with frequency $\omega_{ij}$ plotted on the horizontal axis. (a) 
 shows data for the Bogoliubov modes, while (b) is the corresponding HF calculation.
 Note that in (a) the highest peaks extend far beyond the vertical range, since
 we are most interested in the ``background'' spectrum. The mode oscillation 
 frequency for this radial breathing mode is $\omega_{\rm osc}=2.235 \omega_0$.} 
\end{figure}

We now move on to describing the results of our calculations, where we 
begin with the case of the Bogoliubov method for an isotropic trap 
($\lambda=1$, $\omega_0 \equiv \omega_\perp = 2\pi \times 187\, 
{\rm Hz}$). The specific system being considered is a gas of $^{87}$Rb
atoms with a scattering length $a=5.82 \times 10^{-9}$ m.
The calculation follows closely  that of Guilleumas and Pitaevskii 
(G-P) \cite{guilleumas00}. The results are conveniently displayed as a 
histogram which plots the damping strengths, $\gamma_{ij}$, versus the 
excitation frequency $\omega_{ij}=(E_j-E_i)/\hbar$. 
Fig.~\ref{fig:land-hist}(a) shows such a plot for $N_c=2.5\times 10^5$ 
condensate atoms at a temperature of $k_B T /\mu = 1.5$, where the 
chemical potential takes its Thomas-Fermi value $\mu=29.9 \hbar 
\omega_0$. The spectrum 
features several high peaks, which correspond to transitions between
low-lying excitations with large overlaps (i.e., large $A_{ij}$ matrix 
elements)
together with large population factors, $f_i-f_j$. Following G-P, we shall 
refer to these peaks as ``resonances''. They are not expected to play a 
significant role
in the Landau damping process, unless one happens to lie very close to
the mode frequency $\omega_{\rm osc}$. For this reason we shall ignore these 
transitions when calculating the Landau damping, and instead focus on 
the small amplitude quasi-continuous ``background''.

\begin{figure}[here]
\scalebox{.45}
{\includegraphics{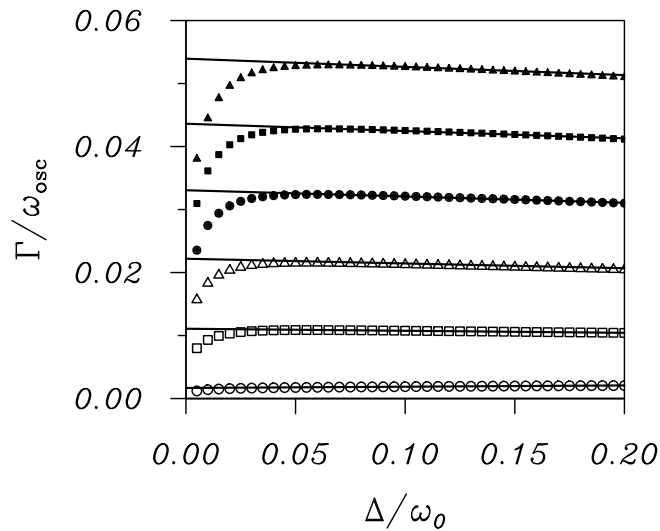}}

\caption{\label{fig:damp-delt}
 Variation of the damping rate (in units of the mode frequency
 $\omega_{\rm osc}$) with
 the Lorentzian width $\Delta$ (in units of the trap frequency), for $\lambda=1$
 and $N_c=5 \times 10^4$. Each curve represents a different temperature, with
 (moving from bottom to top) $k_B T /\mu=$ 0.4, 0.8, 1.2, 1.6, 2.0 and 
 2.4. The
 straight lines are fits to the data between $\Delta/\omega_0=0.05$ and $0.20$,
 where the intercept with the $y$-axis gives the estimated damping rate.}
\end{figure}

\begin{figure}
\scalebox{.5}
{\includegraphics{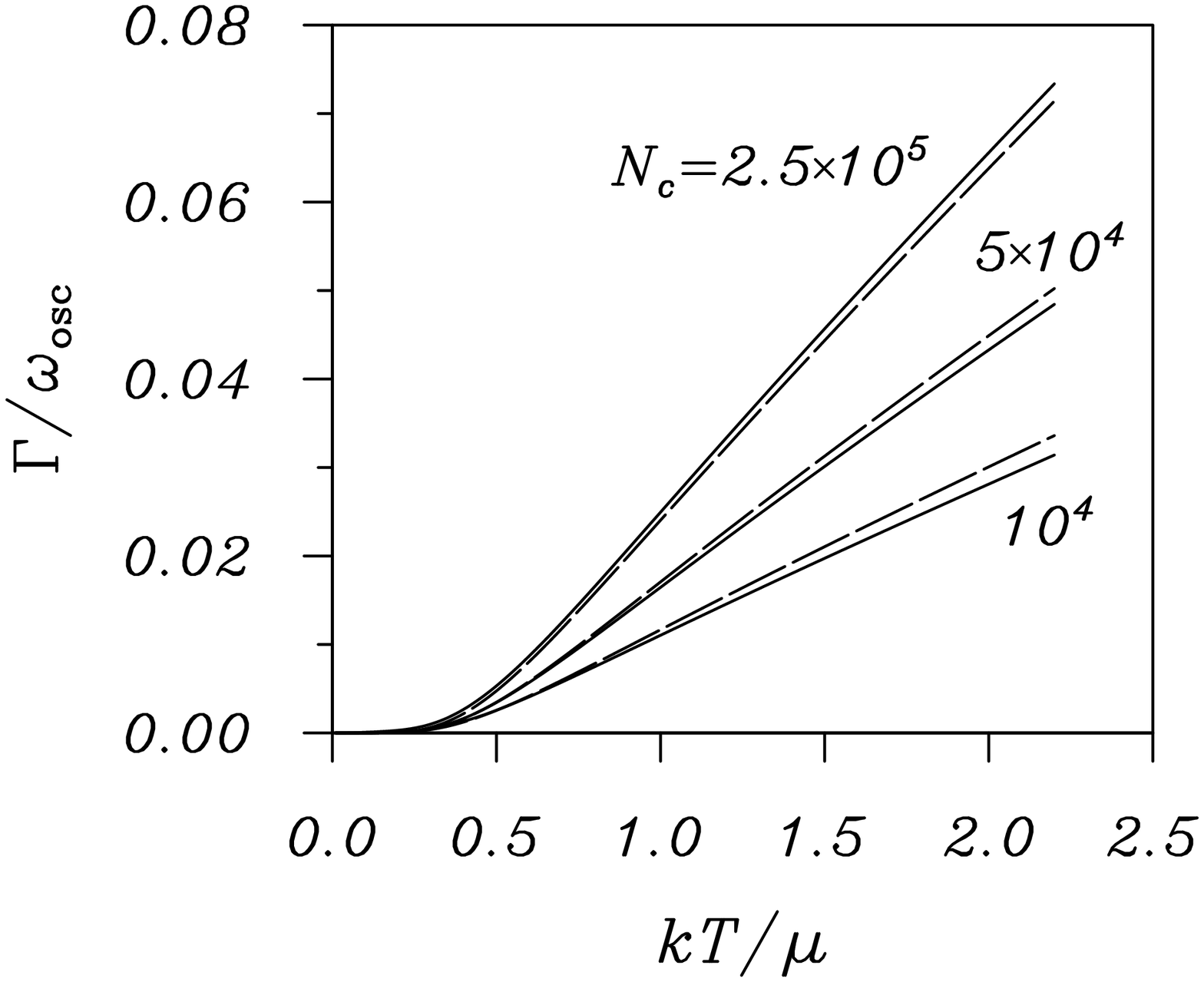}}

\caption{\label{fig:damp-temp}
 Damping rate (in units of $\omega_{\rm osc}$) plotted versus 
 temperature, $k_B T/\mu$,
 where $\mu$ is the Thomas-Fermi chemical potential for each number of 
 condensate atoms, $N_c$. Results from a Bogoliubov calculation are plotted as
 solid lines, while those for the HF approximation are plotted with dashed 
 lines. Results for $N_c = 10^4$, $5\times 10^4$, and $2.5\times 
 10^5$ are shown.}
\end{figure}

A conceptual difficulty arises in the calculation in that
the excitation spectrum in (\ref{dampingrate}) consists of discrete
delta functions. Taken literally, this would imply that the
Landau damping rate would either be zero or infinite depending on the
location of the oscillation frequency $\omega_{\rm osc}$. 
Following \cite{guilleumas00}, this difficulty is overcome by 
replacing each delta function
in the background spectrum by a Lorentzian $\Delta/\{
(2\pi)[(\omega_{ij}-\omega_{\rm osc})^2+\Delta^2/4]\}$. Physically, 
this accounts for the fact that the excitations will themselves have a
finite lifetime in a more rigorous theory. As a result, the Landau
damping rate is
determined by averaging over many peaks weighted according to their 
proximity to the mode frequency. The width factor, $\Delta$, however 
is somewhat arbitrary, and as shown in Fig.~\ref{fig:damp-delt}, 
the result for $\Gamma$ will vary with $\Delta$. Nevertheless, one can 
see that the variation is weak when $\Delta/\omega_0$ lies between 
0.05 and 0.20. We fit the data in this range to a straight line, and 
extrapolate back to $\Delta=0$ to yield an estimate of the Landau 
damping rate. The result as a function of temperature is shown in 
Fig.\ \ref{fig:damp-temp}, for different
numbers of condensate atoms. One sees in the plots the typical 
rapid increase of the rate at low temperatures followed by a nearly
linear temperature dependence at the higher temperatures. Our results
for $N_c = 5\times 10^4$ are in complete agreement with those of
Guilleumas and Pitaevskii~\cite{guilleumas00}.

\begin{figure}
\scalebox{.52}
{\includegraphics{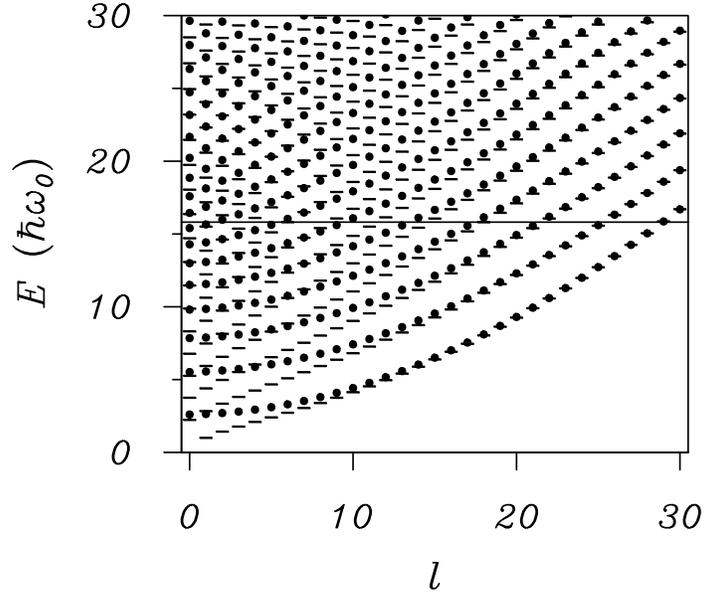}}

\caption{\label{fig:ex-spect}
 Excitation spectrum for Bogoliubov (lines) and HF excitations (bullets), for 
 a spherical condensate with $N_c = 5\times 10^4$. For each mode the energy 
 $E$ and multipolarity $l$ are plotted. The horizontal line at $\mu=15.8 \hbar
 \omega_0$ represents the chemical potential of the condensate.}
\end{figure}

We have repeated the calculation for HF thermal excitations following
the method discussed in Sec.~\ref{HFA}. The spectral density in this case is 
shown in Fig.\ \ref{fig:land-hist}(b), for the same parameters as in 
(a). 
The main difference between the two plots is the absence of the strong 
resonances seen in the Bogoliubov calculation. However, the 
background spectrum is remarkably
similar, especially in the vicinity of the mode frequency $\omega_{\rm
osc}$. Insight 
into why this is happening is provided by comparing the Bogoliubov and
HF excitation frequencies, as plotted in Fig.\ \ref{fig:ex-spect} for 
$N_c=5\times 10^4$. As found previously~\cite{dalfovo99}, the two 
spectra are completely different at low energies and angular momenta, 
but converge at high $E$ and $l$, demonstrating that the
excitations take on a single-particle character \cite{you97,dalfovo97}
in this region. It is these excitations that
are responsible for the majority of the background transitions in Fig.\ 
\ref{fig:land-hist} which explains the similarity between the HF
(calculated in the same way as described above) and Bogoliubov
damping rates plotted in Fig.\ \ref{fig:damp-temp}. 
The excellent agreement with the Bogoliubov results for a wide
range of condensate sizes is in contrast with the uniform Bose gas
\cite{pitaevskii97}, where one finds a significant difference between 
the two
approaches. The reason for this is that the Bogoliubov spectrum for a
uniform gas only approaches the single-particle HF form at temperatures
much larger
than the chemical potential, while in a trapped gas surface excitations with
high multipolarities are also important, even at low temperatures. 
We thus arrive at the important conclusion that the `collective nature' 
of the excitations has little bearing on the determination of 
Landau damping in trapped Bose gases. 

\begin{figure}
\scalebox{.5}
{\includegraphics{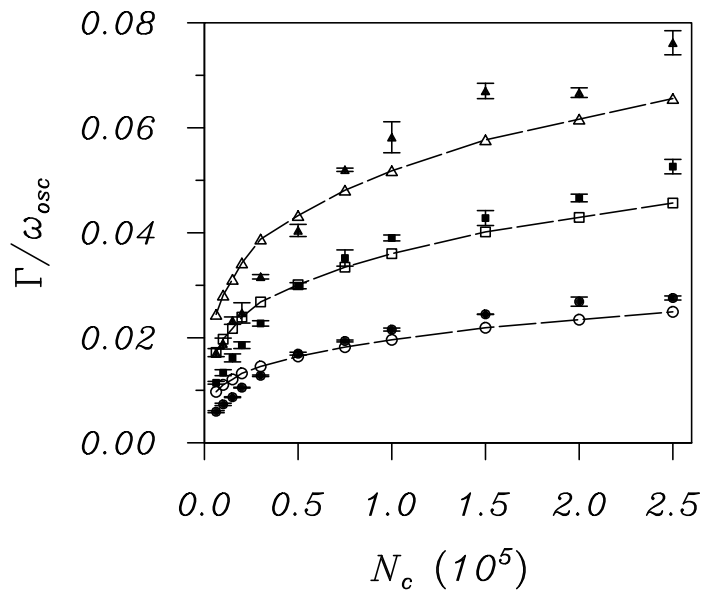}}

\caption{\label{fig:HFB-sim}
 Damping rate (in units of $\omega_{\rm osc}$) against number of condensate 
 atoms, $N_c$ for an isotropic trap. The results from a Bogoliubov
 calculation (open symbols) are compared to those of a semiclassical 
 simulation (solid), for temperatures of $k_B T/\mu =1$ (circles), 
 $k_B T/\mu =1.5$ (squares), and $k_B T/\mu =2$ (triangles). The dashed
 lines through the Bogoliubov data serve as guides to the eye. Each 
 simulation data point is calculated by taking the mean of damping 
 rates extracted from oscillations in the three directions, with the 
 standard deviation giving the error bar.} 
\end{figure}

It is also of interest to compare the Bogoliubov results to those of our
semiclassical simulations based on the ZNG theory (as described in 
Section 2E). We plot
the damping rates for both calculations as a function of the number of 
condensate atoms in Fig.\ \ref{fig:HFB-sim}, for three different 
temperatures. The behaviour seen for both sets of calculations is
similar, with a more rapid rate of increase of the damping rate at 
low $N_c$ followed by a less rapid increase at higher $N_c$.
The data at $N_c=5\times 10^4$ and $N_c=1.5\times 10^5$ 
correspond to those plotted in Fig.\ 8 of Ref.~\cite{jackson02c}, 
where we concluded that the two approaches were in good agreement.
The comparison in Fig.\ \ref{fig:HFB-sim} shows that the agreement 
persists over a wider range of condensate number, although some
systematic differences do exist. The simulation rates at low $N_c$ tend
to be lower than the Bogoliubov damping rate, while at high $N_c$ 
there is a tendency for them to be slightly larger.  These
small differences are presumably due to the different approximations
used in the two sets of calculations. These include (i) the use of HF as
opposed to Bogoliubov excitations in the simulations, (ii) the use of
the semiclassical approximation for the thermal cloud dynamics 
and (iii) the inclusion of the thermal cloud mean field in the 
self-consistent calculation of both the equilibrium and dynamical
properties. Concerning (i), we saw in Fig.~\ref{fig:damp-temp} that
there is little difference between the Bogoliubov and HF results over 
a wide range of temperatures when the excitations are both treated 
quantum mechanically. We therefore do not expect the use of Bogoliubov,
as opposed to HF, quasiparticles to appreciably change the results.

\begin{figure}[here]
\scalebox{.45}
{\includegraphics{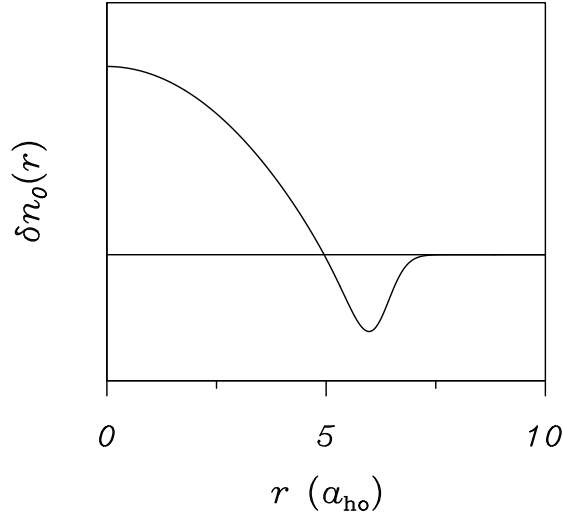}}

\caption{\label{fig:mode-dens}
Mode density (arbitrary units) of the radial breathing mode as a 
function of the radial distance $r$ in units of the harmonic oscillator
length $a_{\rm ho} = \sqrt{\hbar/m\omega_0}$.
The condensate contains $N_c = 10^5$ atoms.}
\end{figure}

The effect of the semiclassical approximation itself, however, is less 
clear. To isolate this effect we make use of the formulation
outlined in Sec. IID.
The calculation is based on Eq.~(\ref{gamma_sc}) with $f(t)$ determined
from (\ref{sum}) by sampling the mode density $\delta n^-(\br)$ along
classical trajectories. To be specific, we again consider the radial
breathing mode in an isotropic trap with frequency $\omega_0 = 2\pi
\times 187$ s$^{-1}$. The mode density being sampled in this case
is shown in Fig.~\ref{fig:mode-dens} for $N_c = 10^5$. The node in the
mode density is of course required in order that the volume
integral of the mode density vanish.

\begin{figure}[here]
\scalebox{.47}
{\includegraphics{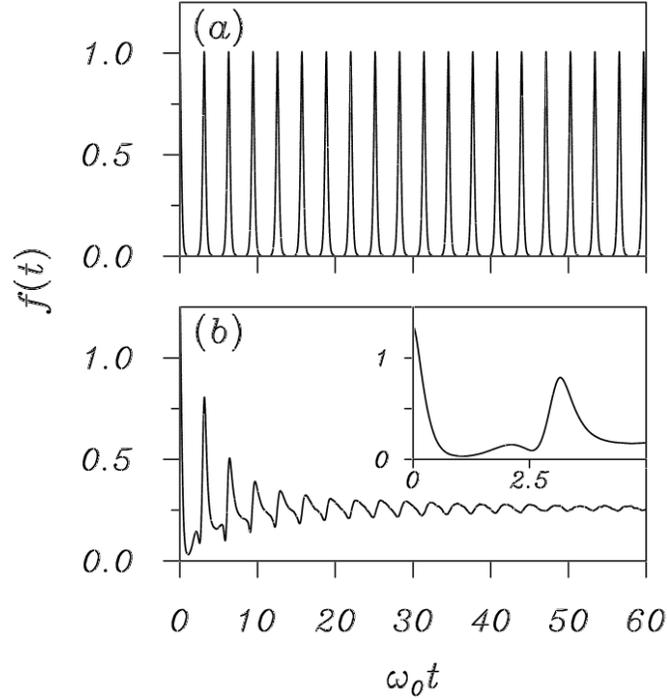}}

\caption{\label{fig:model-calc}
(a) The weighted self-diffusion function $f(t)$ for a cloud of thermal
atoms in an isotropic harmonic potential with frequency $\omega_0 = 2\pi
\times 187$ s$^{-1}$ at a temperature of $k_BT/\mu =1.5$. (b) As in (a)
but with the mean field potential of $N_c = 5\times 10^4$ condensate
atoms included. The vertical scales are in units of $10^{-3}N_c/a_{\rm
ho}^6$.
The Boltzmann distribution used in these calculations is
normalized to the same number of atoms as obtained with the Bose
distribution in Fig.~\ref{fig:Bose_f(t)}.
}
\end{figure}

We consider first a model in which the thermal atoms move in 
a purely harmonic potential, that is, we ignore the mean field of
the condensate.  Furthermore, we assume that the phase space 
density of the atoms is given by a Boltzmann distribution.
With these assumptions, the self-diffusion function $G_s(\br,\br',t)$ 
can be evaluated analytically with the result~\cite{vineyard58}
\beq
G_s(\br,\br',t) = \tilde N \left ( {\beta m \omega_0^2 \over 2\pi |\sin \omega_0 t|}
\right )^3 \exp{\left ( -{\beta m \omega_0^2 (r^2 + r'^2 - 
2\br\cdot\br' \cos \omega_0 t) \over 2\sin^2\omega_0 t }\right )}\,.
\label{G_s_harmonic}
\eeq
This function is periodic with period $T_0 = 2\pi/\omega_0$: it recovers
its $t=0$ value $G_s(\br,\br',0) = \delta(\br-\br') \tilde n_0(\br)$ at
the times $t_n = nT_0$. This is a special property of purely harmonic 
confinement. When substituted into (\ref{f(t)}), the resulting 
function $f(t)$ has period $T_0/2$ since the mode density $\delta
n^-(\br)$ has inversion symmetry. We show in 
Fig.~\ref{fig:model-calc}(a) $f(t)$ as calculated from
(\ref{sum}) by following the trajectories of an ensemble of test 
particles at a temperature $k_BT/ \mu =1.5$.  The
figure confirms the behaviour expected on the basis of the analytic form
of the self-diffusion function in (\ref{G_s_harmonic}). 
In Fig.~\ref{fig:model-calc}(b) we show a similar calculation of $f(t)$
but now with the condensate mean field included. The distribution of
thermal atoms is still Boltzmann-like. The strict periodicity seen in 
Fig.~\ref{fig:model-calc}(a) is washed out as a result of the motion of 
the
thermal atoms in the nonharmonic confining potential, $V_{\rm th}(\br) =
V_{\rm ext}(\br) + 2gn_c(\br)$, although there are
clearly vestiges of the dominant $2\omega_0$ frequency of the purely
harmonic case. The mean field of the condensate disrupts the 
periodicity of the particle orbits
that occurs for harmonic confinement and as a result,
$f(t)$ saturates at long times to a constant limiting value. 
The inset of Fig.~\ref{fig:model-calc}(b) shows $f(t)$ on a smaller time
scale; the behaviour seen is consistent with $f(t)$ being an even 
function of time.

\begin{figure}[here]
\scalebox{.45}
{\includegraphics{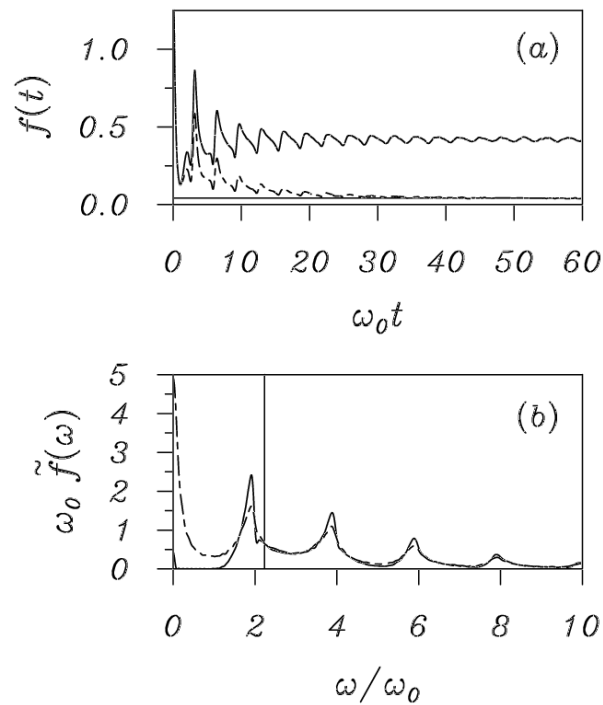}}

\caption{\label{fig:Bose_f(t)}
(a) As in Fig.~\ref{fig:model-calc}, but with the Boltzmann distribution
replaced by the Bose distribution (solid line). The dashed curve shows
the effect of including collisions between the thermal atoms. The
horizontal asymptote at $f_{\rm eq}$, Eq.~(\ref{f_eq}), is shown. (b) 
The Fourier
transform of the weighted self-diffusion function for the two curves in
(a): no collisions (solid), with collisions (dashed). Both vertical
scales have the same units as in Fig.~\ref{fig:model-calc}. The vertical
line indicates the mode frequency at $\omega_{\rm osc}
= 2.235\omega_0$.
}
\end{figure}

In Fig.~\ref{fig:Bose_f(t)}(a) we show $f(t)$ for the same conditions 
as in Fig.~\ref{fig:model-calc}(b), but
with the Boltzmann distribution replaced by the Bose distribution. The
behaviour of $f(t)$ is qualitatively very similar; the main difference
is the limiting long-time value which can be attributed to the different
thermal distributions in the two simulations. It is worth commenting on
what this asymptotic value is due to. If the position of an atom at long
times is uncorrelated with its starting point, one might expect the
self-diffusion function to tend towards $\tilde n_0(\br)\tilde
n_0(\br')/\tilde N$. This would yield an `equilibrium' value of $f(t)$
of
\beq
f_{\rm eq} \equiv {1\over \tilde N} \left | \int d\br\,
\delta n^{-}(\br)\tilde n_0(\br) \right |^2\,.
\label{f_eq}
\eeq
We find, however, that $f(t)$ in our simulations does not tend to this
limit. The reason for this is that long-time correlations persist since
a given atom retains its initial energy in the course of the dynamical 
evolution. Even if the nonharmonic perturbation were to lead to
an ergodic distribution on a given
energy surface, the equilibrium limiting form can only arise if 
equilibrating collisions between thermal atoms take place. This was
confirmed by performing a simulation in which collisions are included
following the methods discussed in Ref.~\cite{jackson02c}. The dashed 
curve in Fig.~\ref{fig:Bose_f(t)}(a) shows
the result of this calculation and we now indeed find that $f(t)
\to f_{\rm eq}$ as $t \to \infty$.

In Fig.~\ref{fig:Bose_f(t)}(b) we show the
Fourier transform of $f(t)$. It has peaks at frequencies close to
multiples of $2\omega_0$ as would be expected in view of
Fig.~\ref{fig:Bose_f(t)}(a). Less
obvious is the fact that $\tilde f(\omega)$ appears to be
positive definite. This behaviour, however, is necessary since $\tilde
f(\omega)$ is just the semiclassical limit of the spectral density in 
(\ref{rate2}). It is therefore reassuring that this property emerges 
from our simulations. The dashed curve in  Fig.~\ref{fig:Bose_f(t)}(b)
is the corresponding Fourier transform when collisions are included. The
main difference between the two curves occurs at low frequencies where
the collisional curve has a Lorentzian peak due to the relaxation of
$f(t)$ to $f_{\rm eq}$ on a time scale of $\omega_0 t \sim 10$.

\begin{figure}[here]
\scalebox{.45}
{\includegraphics{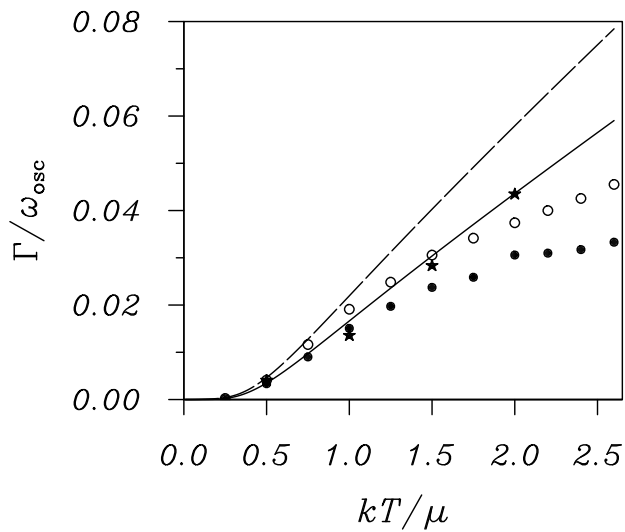}}

\caption{\label{fig:sc_damping}
A comparison of the Bogoliubov damping rates (lines) with the 
semiclassical HF approximation (points): $N_c = 5\times 10^4$, solid
line and filled points; $N_c = 1.5 \times 10^5$, dashed line and
unfilled points. The stars represent the result of a semiclassical
simulation for $N_c = 5\times 10^4$ as described in the text. 
}
\end{figure}
\begin{figure}[here]
\scalebox{.45}
{\includegraphics{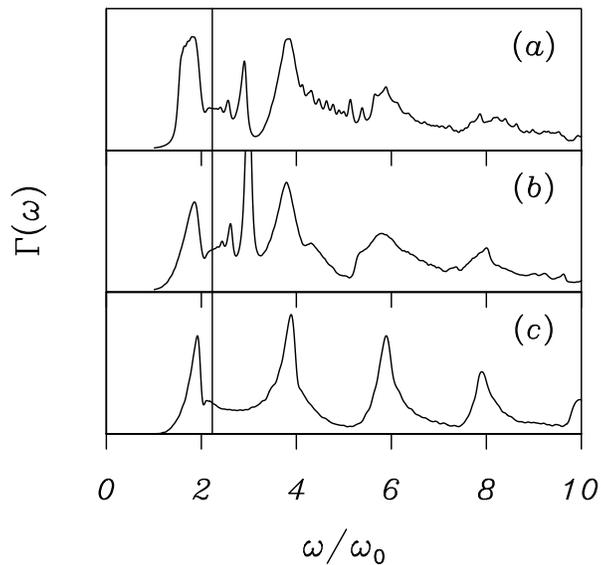}}

\caption{\label{fig:comparison}
Comparison of the (a) Bogoliubov, (b) HF, and (c) semiclassical HF
spectral densities (all having the same arbitrary scale) as discussed in
the text. The calculations are for
a temperature of $k_BT/\mu =1.5$ and $N_c = 5\times 10^4$ condensate
atoms. The vertical line at $\omega = 2.235\omega_0$ indicates the
position of the radial breathing mode.
}
\end{figure}

The semiclassical Landau damping rate is determined by the value of 
$\tilde f(\omega)$ at $\omega = \omega_{\rm osc}$  for the radial
breathing mode. This frequency is indicated 
by the vertical line in Fig.~\ref{fig:Bose_f(t)}(b) and it
is clear that collisions are not important in this case. 
The Landau damping rates calculated
on the basis of (\ref{gamma_sc}) are plotted as a function of 
temperature in Fig.~\ref{fig:sc_damping}
which also displays the Bogoliubov results for comparison. The two sets
of calculations are qualitatively similar and show a similar dependence
on the condensate number $N_c$. However, the semiclassical results
underestimate the Bogoliubov results at higher temperatures. 

To understand these differences we show in
Fig.~\ref{fig:comparison} excitation spectral densities for the three
approximations on a much larger frequency range. 
For the Bogoliubov and HF calculations, the spectral
densities are given by (\ref{dampingrate}), with $\omega_{\rm osc}$
replaced by an arbitrary frequency $\omega$, and convolved with a
Lorentzian of width $\Delta = 0.1\omega_0$. This representation is more
informative than the histograms in Fig.~\ref{fig:land-hist} since it
reveals the total spectral weight as a function of frequency.
The corresponding semiclassical spectral density follows from 
(\ref{gamma_sc}) and is given by
\beq
\Gamma_{\rm sc}(\omega)={2g^2\over\hbar k_B T}\omega\tilde f(\omega)\,.
\eeq
It is clear from this figure that the spectral densities in all three
approximations are qualitatively similar, with a series of broad peaks
near multiples of $2\omega_0$. However, both the Bogoliubov and HF
calculations have a higher spectral density in the range $2\omega_0 <
\omega < 4\omega_0$ than the semiclassical result, which accounts for
the smaller semiclassical damping rates shown in
Fig.~\ref{fig:sc_damping}. The larger quantum spectral
densities are associated with low-energy states for which the
zero-point energy and the effects of barrier penetration are
more important. These quantum effects are of course missed in the 
semiclassical HF calculation. However, the overall similarity of the
curves in Fig.~\ref{fig:comparison} indicates that the semiclassical HF
approximation is providing a good qualitative description of the thermal
cloud excitations.

In order to gain more insight into these results,
we have performed another simulation which incorporates the same physics
as the semiclassical HF approximation formulated in
Sec.~\ref{SCHF}.  As in the full simulations in Sec. \ref{simulations}, 
we evolve the condensate
using the time-dependent GP equation but do not include the mean field
of the thermal cloud acting on the condensate. The condensate thus
oscillates with a fixed amplitude giving rise to a harmonic perturbation
of the thermal cloud which itself starts off as an equilibrium 
distribution and evolves classically in time in
the presence of the dynamic mean field of the condensate. These are the
same ingredients entering the HF perturbation theory calculation.
We then monitor the thermal cloud energy $E_{th}(t) = \sum_i \left [ 
p_i^2(t)/2m + V_{th}(\br_i(t)) \right ]$, where $V_{th}(\br)$ is the
equilibrium effective potential acting on the thermal cloud. As a result
of the work done by the condensate on the thermal cloud, the latter
experiences a secular increase in energy which we can identify with the
time-averaged rate of energy transfer in (\ref{rate2}). Some points
obtained in this way are shown in Fig.~\ref{fig:sc_damping}. 
These points agree quite well
with the Bogoliubov rates, and by the same token, with the full
simulations as shown in Fig.~\ref{fig:HFB-sim}. Thus, both types of
simulations are mutually consistent but deviate from the
results based on the semiclassical HF expression in (\ref{gamma_sc}).
Apparently, following the dynamics of the thermal cloud in the presence
of an oscillating condensate is not entirely equivalent to the
evaluation of the equilibrium correlation function in the semiclassical
HF expression.
At the present time we do not have a satisfactory explanation for the
observed difference. We speculate that it may be related to the 
nonequilibrium nature of the thermal cloud distribution in the 
simulations, but we have not been able to find a way to check this.

\subsection{Anisotropic Traps}

We now describe the effect of introducing an anisotropy  ($\lambda<1$)
such that the condensate takes on a prolate geometry. The anisotropy 
lifts the degeneracy with respect to the $m$ quantum number which
necessitates the inclusion of a much larger number of transitions in
our Bogoliubov calculations. As a result, calculations for
$N_c > 10^4$ are numerically prohibitive.
The same is true for large anisotropies, so we are limited 
here to $0.3 < \lambda \le 1$. This range, however, is sufficiently
large to reveal some general trends.

\begin{figure}[here]
\scalebox{.5}
{\includegraphics{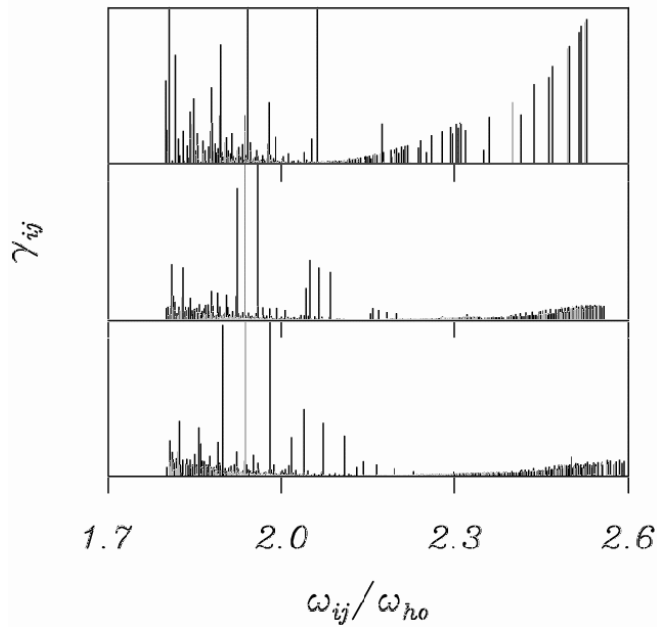}}

\caption{\label{fig:spec-evo} Histogram of Landau damping strength,
 $\gamma_{ij}$, versus excitation frequency $\omega_{ij}$ for the 
 $m=0$ breathing mode in a condensate with $N_c=10^4$ atoms 
 at a temperature of $k_B T/\mu=1.5$. Each plot is for a different 
 trap anisotropy: from top to bottom, $\lambda=$ 1, 0.95, and 0.9.
 The geometric mean of the trap frequency in all cases is 
 $\omega_{\rm ho}=2\pi \times 187\, {\rm Hz}$.}
\end{figure}

The effect of imposing a slight anisotropy on the excitation spectrum is
shown in Fig.\ \ref{fig:spec-evo} (for $N_c = 10^4$ and with the 
geometric mean of the trap frequency $\omega_{\rm ho}=\omega_\perp 
\lambda^{1/3}$ kept fixed at $2\pi \times 187\, {\rm Hz}$). 
One sees that some of the larger resonances, 
corresponding to transitions between the lower-energy excitations, are
each split into a series of shorter peaks. This is most clearly 
illustrated for the large peak at $\omega_{ij}/\omega_{\rm ho}=1.94$, 
which is associate with a pair of low-lying $l=1$ modes. 
The anisotropy splits this peak into three since
the $m=-1$, 0 and 1 modes are no longer degenerate. As the anisotropy 
increases the peaks continue to separate, as shown by the bottom panel 
in Fig.\ \ref{fig:spec-evo} for $\lambda=0.9$.     

\begin{figure}[here]
\scalebox{.45}
{\includegraphics{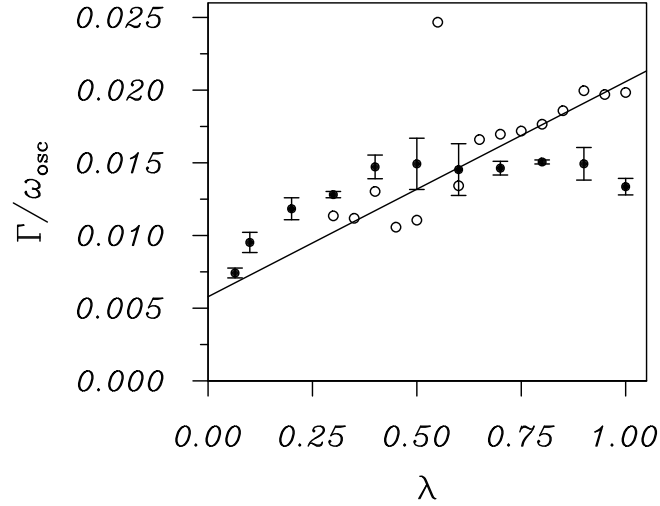}}

\caption{\label{fig:damp-ani} Damping rate of the monopole mode (in units of 
 the mode frequency, $\omega_{\rm osc}$) in an axisymmetric trap, as a function of
 the anisotropy, $\lambda$. The open circles and closed circles represent the 
 Bogoliubov and simulation results respectively. The solid line is a straight
 line fit to the Bogoliubov data, neglecting the outlying data point at 
 $\lambda=0.55$.} 
\end{figure}

\begin{figure}[here]
\scalebox{.5}
{\includegraphics{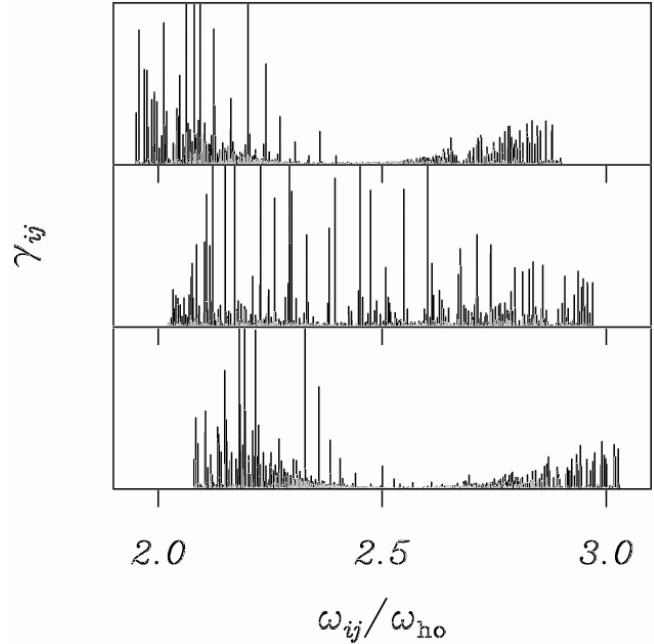}}

\caption{\label{fig:spec-evo2} Same as Fig.\ \ref{fig:spec-evo} but for 
 anisotropies of (from top to bottom) $\lambda=$ 0.6, 0.55 and 0.5, 
 showing the origin of the anomalously high damping rate at 
 $\lambda=0.55$. The shift of the spectrum towards the right is due to 
 the ratio $\omega_\perp/\omega_{\rm ho}$
 increasing as $\lambda$ increases.} 
\end{figure}

An obvious consequence of this is that the spectrum is now much denser, but one
can extract the damping rate in much the same way as described above for 
isotropic traps. The only difference is that we no longer ignore the larger 
peaks (resonances) in our calculation, since these are more numerous 
and less distinguishable from the background. There is therefore no
simple criterion that can be used to remove them.
The damping rate for each $\lambda$ is plotted in Fig.\ 
\ref{fig:damp-ani}, where one sees a downward trend as the anisotropy 
increases towards a cigar-shaped geometry (decreasing $\lambda$). 
The anomalously high damping rate at $\lambda=0.55$
is a consequence of a mixing of the radial breathing mode 
with another $m=0$ mode which happens to possess a similar frequency
at this anisotropy. We show the effect of this mode crossing on the 
spectrum by scanning through $\lambda=0.55$ in Fig.~\ref{fig:spec-evo2}.
Many more high resonances are seen at $\lambda=0.55$ than to either side
which in turn leads to the much higher damping rate shown in 
Fig.\ \ref{fig:damp-ani}. We have here a case in which the `resonant
modes' do in fact contribute to the Landau damping. We also believe
that the scatter in the Bogoliubov data seen in
Fig.~\ref{fig:spec-evo2} can be accounted for in terms of resonant 
peaks approaching or receding from the vicinity of the breathing mode
frequency.

The straight line in Fig.\ \ref{fig:damp-ani} 
is a least-squared fit to the Bogoliubov data (excluding the high-lying 
point at $\lambda=0.55$). Extrapolating the straight line fit to 
$\lambda=0$ suggests a damping rate of $\lambda/\omega_{\rm osc} 
\simeq 0.006$ in this limit, which is around a factor of 4 
smaller than in the spherical case. For comparison, we have also
plotted in Fig.\ \ref{fig:damp-ani} the results of semiclassical ZNG 
simulations. These calculations are again
carried out for the same condensate number ($N_c=10^4$) and temperature
($k_B T/\mu = 1.5$). We note from
Fig.~\ref{fig:HFB-sim} that the simulation damping rate at $\lambda =1$
lies below the Bogoliubov calculation for this low number of condensate
atoms. This seems to impart a slightly different $\lambda$ dependence 
as compared to the Bogoliubov results but we still see a downward trend
for $\lambda < 0.5$. In particular, there is a rapid 
decrease in the simulation damping rate for $\lambda \to 0$.

These results shed some light on differences found in previously
reported calculations~\cite{jackson02b,guilleumas02}.
Guilleumas and Pitaevskii~\cite{guilleumas02} performed
a Bogoliubov calculation for a cylindrical condensate (corresponding 
to $\lambda=0$) and found a Landau damping about
two order of magnitudes smaller than the typical values in 
Fig.~\ref{fig:HFB-sim}, and an order of magnitude {\it smaller} than the
damping rates observed experimentally for $\lambda =
0.065$~\cite{chevy02}. On the other hand, we found~\cite{jackson02b}
a conventional Landau damping rate from ZNG simulations
(as discussed in
this paper) that was consistent with Fig.~\ref{fig:HFB-sim}, but an
order of magnitude {\it larger} than experiment. 
The rapid variation of the simulation results in Fig.~\ref{fig:damp-ani}
for small $\lambda$ is a possible resolution of these different 
theoretical predictions. If the damping rate continues to drop rapidly
for $\lambda \to 0$, one might find a result consistent with that of
Guilleumas and Pitaevskii~\cite{guilleumas02}. Unfortunately,
we cannot push our calculations to lower $\lambda$
in order to confirm this. It is nevertheless clear from all of our
calculations that the Landau damping in the cylindrical geometry appears
to be an anomalous limit.

As a final point, we should emphasize that the conventional Landau
damping rate discussed here fails to account for the recent observations
of the damping of the transverse breathing mode in elongated 
condensates~\cite{chevy02}. This
could only be achieved by including the full dynamics of the thermal
cloud in the ZNG simulation~\cite{jackson02b}. Conventional Landau
damping deals with the oscillation of the condensate in an otherwise
equilibrium thermal cloud; the condensate does work on the thermal cloud
and loses energy to it. However, if the thermal cloud is itself set into
motion as is the case in the experiments~\cite{chevy02}, a
change in the damping rate is to be expected. The experimental
observations~\cite{chevy02} can only be accounted for if this effect is
taken into account~\cite{jackson02b}.

\section{Conclusions}

In this paper we have studied Landau damping of Bose-Einstein 
condensates in isotropic and anisotropic traps and have compared 
a variety of methods for calculating the damping. 
Calculations based
on perturbation theory, involving sums over transitions between excited states,
show excellent agreement between a Bogoliubov treatment of the
excitations and the Hartree-Fock
approximation over a wide range of temperatures and numbers of atoms. 
These results demonstrate that Landau damping in trapped Bose-condensed
gases is essentially a `single-particle' phenomenon in that the
collective nature of the excitations is unimportant. 
A semiclassical limit of the HF approximation has also been developed,
leading to a novel expression for the damping rate in terms of a
classical correlation function. A detailed comparison with the quantal
HF results has shown that the excitation spectra in the two cases are
qualitatively similar but differ at a quantitative level. In particular,
the semiclassical results tend to underestimate the quantum results at
higher temperatures. We also explored damping by means of 
$N$-body simulations (as discussed in Ref.\ \cite{jackson02c}) in 
two complementary ways. In the first, the condensate wavefunction is
evolved using the time-dependent GP equation and its interaction with 
the thermal cloud leads to a decay of the condensate 
oscillation amplitude. In the second, the condensate oscillates with a
fixed amplitude and the rate of increase of the thermal cloud energy is
determined. Both methods give very similar results which agree with the
quantum perturbation results over a wide range of temperatures and 
condensate sizes, reinforcing the conclusions of Ref. \cite{jackson02c}.
However, all of these results differ somewhat (see
Fig.~\ref{fig:sc_damping}) from the semiclassical HF formulation. At
present we have no explanation for these differences.

We have also studied in detail the damping in cigar-shaped traps using 
the Bogoliubov approximation and semiclassical simulations. These 
results show a decreasing trend in 
the damping as the trap becomes increasingly anisotropic. Over the range
of $\lambda$ in both sets of calculations,
the damping decreases by about a factor of two, with the semiclassical
simulations indicating a more rapid decrease for $\lambda \to 0$. This
behaviour may reconcile the differences in the results obtained for 
highly elongated condensates~\cite{jackson02b} and those
obtained~\cite{guilleumas02} in the cylindrical limit ($\lambda = 0$).

\section{Acknowledgements}

We acknowledge the use of the HPCVL computer facility at Queen's University,
together with financial support from the Natural Sciences and Engineering
Research Council of Canada. 
   
\bigskip
{\noindent $^\dagger$Present Address: Dipartimento di Fisica, 
Universit\`{a} di Trento and BEC-INFM, I-38050 Povo, Italy.}


\begin{references}

\bibitem{jin97} D. S. Jin, M. R. Matthews, J. R. Ensher, C. E. Wieman, and
 E. A. Cornell, Phys. Rev. Lett. {\bf 78}, 764 (1997).
\bibitem{stamperkurn98} D. M. Stamper-Kurn, H.-J. Miesner, S. Inouye, 
 M. R. Andrews, and W. Ketterle, Phys. Rev. Lett. {\bf 81}, 500 (1998).
\bibitem{marago01} O. Marag\`{o}, G. Hechenblaikner, E. Hodby, and 
 C. J. Foot, Phys. Rev. Lett. {\bf 86}, 3938 (2001).
\bibitem{chevy02} F. Chevy, V. Bretin, P. Rosenbusch, K. W. Madison, and 
 J. Dalibard, Phys. Rev. Lett. {\bf 88}, 250402 (2002). 
\bibitem{zaremba99} E. Zaremba, T. Nikuni, and A. Griffin, J. Low Temp.
 Phys. {\bf 116}, 277 (1999). 
\bibitem{jackson02c} B. Jackson and E. Zaremba, Phys. Rev. A {\bf 66}, 
033606 (2002).
\bibitem{jackson01} B. Jackson and E. Zaremba, Phys. Rev. Lett. {\bf 87},
 100404 (2001).
\bibitem{jackson02a} B. Jackson and E. Zaremba, Phys. Rev. Lett. {\bf 88},
 180402 (2002).
\bibitem{jackson02b} B. Jackson and E. Zaremba, Phys. Rev. Lett. {\bf 89},
 150402 (2002).
\bibitem{jackson02d} B. Jackson and E. Zaremba, Laser Phys. {\bf 12},
 93 (2002).
\bibitem{hohenberg65} P. C. Hohenberg and P. C. Martin, Ann. Phy. {\bf 34},
 291 (1965).
\bibitem{szepfalusy74} P. Sz\'{e}pfalusy and I. Kondor, Ann. Phys. {\bf 82},
 1 (1974).
\bibitem{liu97} W. V. Liu, Phys. Rev. Lett. {\bf 79}, 4056 (1997).
\bibitem{pitaevskii97} L. P. Pitaevskii and S. Stringari, Phys. Lett.
 {\bf 235}, 398 (1997).
\bibitem{fedichev98} P. O. Fedichev, G. V. Shlyapnikov, and J. T. M. 
 Walraven, Phys. Rev. Lett. {\bf 80}, 2269 (1998).
\bibitem{giorgini98} S. Giorgini, Phys. Rev. A {\bf 57}, 2949 (1998).
\bibitem{giorgini00} S. Giorgini, Phys. Rev. A {\bf 61}, 063615 (2000).
\bibitem{reidl00} J. Reidl, A. Csord\'{a}s, R.Graham, and 
 P. Sz\'{e}pfalusy, Phys. Rev. A {\bf 61}, 043606 (2000).
\bibitem{guilleumas00} M. Guilleumas and L. P. Pitaevskii, Phys. Rev. A 
 {\bf 61}, 013602 (2000).
\bibitem{das01} K. Das and T. Bergeman, Phys. Rev. A {\bf 64}, 013613 (2001).
\bibitem{guilleumas02} M. Guilleumas and L. P. Pitaevskii, preprint
 cond-mat/0208047.
\bibitem{hutchinson98} D. A. W. Hutchinson and E. Zaremba, Phys. Rev. A 
 {\bf 57}, 1280 (1998).
\bibitem{messiah66} A. Messiah, {\it Quantum Mechanics} (North-Holland, 
 Amsterdam, 1966), Vol.\ 2.
\bibitem{kubo57}
R. Kubo, J. Phys. Soc. Jap. {\bf 12}, 570 (1957).
\bibitem{vanhove54}
The self-diffusion function as conventionally defined (see, for example,
L. Van Hove, Phys. Rev. {\bf 95}, 249 (1954)) is given by
$\bar G_s(\br,t)\equiv \tilde N^{-1}\int d\br' G_s(\br+\br',\br',t)$.
\bibitem{dalfovo99} F. Dalfovo, S. Giorgini, L. P. Pitaevskii, and
 S. Stringari, Rev. Mod. Phys. {\bf 71}, 463 (1999).
\bibitem{you97} L. You, W. Hoston, and M. Lewenstein, Phys. Rev. A {\bf 55},
 R1581 (1997).
\bibitem{dalfovo97} F. Dalfovo, S. Giorgini, M. Guilleumas, L. Pitaevskii, and
 S. Stringari, Phys. Rev. A {\bf 56}, 3840 (1997).
\bibitem{vineyard58}
The corresponding quantity $\bar G_s(\br,t)$ as defined in
Ref.~\cite{vanhove54} was calculated by
G. H. Vineyard, Phys. Rev. {\bf 110}, 999 (1958).







\end{references}
\end{document}